\documentclass[12pt]{article}
\pdfoutput=1
\usepackage{amsmath}
\usepackage{amssymb}
\usepackage{graphicx}
\usepackage{subcaption}
\usepackage{url}
\usepackage{arydshln}
\usepackage[sort&compress, numbers, merge]{natbib}

\usepackage{multirow}

\def\GEV{\mathrm{GeV}}

\usepackage[colorlinks=true, linkcolor=blue, citecolor=blue,
urlcolor=black]{hyperref}

\begin{document}

\begin{titlepage}
\begin{center}

\hfill DESY 15-130 \\
\hfill UT-15-27 \\
\vspace{2mm}
\hfill \today

\vspace{3.0cm}
{\large\bf Testing ATLAS Diboson Excess with Dark Matter Searches at LHC }

\vspace{1.0cm}
{\bf Seng Pei Liew}$^{(a)}$ and 
{\bf Satoshi Shirai}$^{(b)}$ 
\vspace{1.0cm}

{\it
$^{(a)}${Department of Physics, University of Tokyo, Tokyo 113-0033, Japan}\\
$^{(b)}${Deutsches Elektronen-Synchrotron (DESY), 22607 Hamburg, Germany} \\
}

\vspace{1cm}
\abstract{
The ATLAS collaboration has recently reported a $2.6 \sigma$ excess in the search for a heavy resonance decaying into a pair of weak gauge bosons. Only fully hadronic final states are being looked for in the analysis. If the observed excess really originates from the gauge bosons' decays, other decay modes of the gauge bosons would inevitably leave a trace on other exotic searches.
In this paper, we propose the use of the $Z$ boson decay into a pair of neutrinos to test the excess.
This decay leads to a very large missing energy and can be probed with conventional dark matter searches at the LHC.
We discuss the current constraints from the dark matter searches and the prospects.
We find that optimizing these searches may give a very robust probe of the resonance, even with the currently available data of the 8 TeV LHC.
}
\end{center}
\end{titlepage}

\setcounter{footnote}{0}

\section{Introduction}
The ATLAS collaboration has recently reported a search for diboson resonances with $W$ and $Z$ boson-tagged jets at 8 TeV at the Large Hadron Collider (LHC)~\cite{Aad:2015owa}. The most prominent excess is seen at resonance mass approximately 2 TeV in the $WZ$ final-state channel with a significance of $3.4\sigma$. As only fully hadronic final states are considered, the gauge bosons are not well distinguished in the analysis, and the same excess can also be interpreted as $WW$ ($2.6\sigma$) or $ZZ$ ($2.9\sigma$) resonance.

The excess is not statistically significant yet and a similar search by the CMS collaboration has seen no clear excess~\cite{Khachatryan:2014hpa}.
It is too early to conclude that this is a signal of physics beyond the standard model (SM).
Therefore, it is essential to test the excess from many different perspectives.
The purpose of this paper is to investigate as model-independent as possible the consistency of the observed excess of boson-tagged jets with other LHC searches. 

The direct interpretation of the excess in terms of physics beyond the SM would be a new process
$P+P' \to X \to \phi(\to 2j)+\phi' (\to 2j)$, where $P^{(')}$ is a parton,
$X$ a new heavy resonance of mass around 2 TeV, $\phi^{(')}$, a resonance of mass around 100 GeV (We note that the observed excess does not necessarily imply that $\phi^{(')}$ is the SM weak gauge boson), and $j$ a quark or gluon jet. One expects collider signals other than the boson-tagged jet resonances once the above process is assumed, the most promising one being the inverse process, $P+P' \to X \to P+P'$, which induces a resonance from two QCD jets (dijet resonance).
However, the SM background of the dijet resonance with non boson-tagged jets are much larger than the case of the boson-tagged jets~\cite{1407.1376,1501.04198}, and this process is not so significant at probing the new resonance.

If we further assume that the $\phi$'s are weak gauge bosons, signals from other exotic searches in addition to analyses mentioned above are expected, as the non-hadronic decays of the gauge bosons will lead to a variety of final states.
For instance, searches of $X\to W+V(\equiv W/Z)$ followed by $W\to\ell\nu$ and $V \to 2j$~\cite{Khachatryan:2014gha, Aad:2015ufa} provide a strong constraint on the 2 TeV resonance.
Although the leptonic branching fraction of the $Z$ boson is small,  the relatively smaller number of SM background events enables one to impose a stringent constraint with the $X\to Z(\to \ell^+\ell^-)+V(\to 2j)$ channel \cite{Khachatryan:2014gha, Aad:2014xka}.
Such consistency check in the context of the ATLAS diboson excess has been studied in Refs.~\cite{Franzosi:2015zra,Cheung:2015nha,Thamm:2015csa,Brehmer:2015cia}. There are also studies considering bounds from the decay of the new resonance into heavy quarks or the Higgs boson, such as~\cite{Hisano:2015gna,Dobrescu:2015qna,Abe:2015uaa}, but it should be noted that such considerations are model-dependent.

In this paper, we focus on the decay $Z\to \bar{\nu}\nu$, which is the secondary decay mode of the $Z$ boson.
The branching fraction of $X \to Z(\to \bar{\nu}\nu)+ V(\to 2j)$ is relatively large, and a significant number of signal events can be expected.
The signal is a (boson-tagged) jet plus large missing transverse energy (MET).
This channel is not well discussed in the context of the resonance search.
However, such signal can be probed with the available LHC dark matter searches of large MET plus a high transverse momentum jet with/without boson tagging \cite{Aad:2013oja,Aad:2015zva}.

Using the reported LHC dark matter searches, we show that the null results of these searches are constraining the region favored by the ATLAS diboson excess. In addition, we show that by optimizing the MET cuts of the relevant analyses, most of the region can be tested even only with the 8 TeV data. Furthermore, we study the projections of these channels at the next run of the LHC. It is shown that these channels can be a complementary and potentially powerful probe of the diboson resonance.    

The rest of the paper is organized as follows.
In Sec. \ref{sec:DM}, we briefly review the status of the dark matter searches at the LHC.
We recast constraints from dark matter searches on the observed excess in Sec. \ref{sec:RECAST}.
In Sec. \ref{sec:13}, we discuss the prospect of these dark matter searches for the run 2 of the LHC.
Summary and discussion are given in Sec.~\ref{sec:summary}.

\section{LHC Dark Matter Search} \label{sec:DM}
The search for collider signatures of dark matter is an essential part of the LHC physics program. Presumably dark matter particles are produced recoiling against a hard parton inside the detector, resulting in a single-parton final state accompanied with large MET. These final-state topologies are well-suited to our purposes of constraining diboson channels of which one of the gauge boson decays into a pair of neutrinos (mimicking dark matter in practice), while the other decays into particles visible to the detector. For the case where the gauge boson decays leptonically, i.e., $W\to \ell \nu$ or $Z\to {\ell}^+  \ell^- $, we find that the constraints are less sensitive than the corresponding hadronic channels. We hence hereafter focus on the dark matter searches in the hadronic channels.   

We study the channel $X\to Z(\to \bar{\nu} \nu) + V (\to 2j)$ and $m_X \simeq 2$ TeV.
In this channel, $V$ is highly boosted and the produced two quarks are almost collinear.
Such quarks are detected as a single jet with the conventional jet reconstruction algorithm.
By analyzing the substructure of the jet, one can in principle identify that such jets are boson-originated (boson-tagged jets).

\subsubsection*{Mono-jet Search}
Both ATLAS and CMS collaborations have reported searches for a single jet accompanied with large MET (mono-jet) at 8 TeV~\cite{Aad:2015zva,Khachatryan:2014rra}. We utilize only the ATLAS analysis because a more stringent MET cut is imposed, which results in a better sensitivity towards testing the 2 TeV resonance.

Let us summarize the major features of the event selection for the ATLAS analysis. The analysis requires at least a jet with transverse momentum $p_T > 120~\GEV$ and no leptons in the final state. Jets are defined by the anti-$k_T$ jet algorithm, with the radius parameter $R=0.4$~\cite{0802.1189}. Several signal regions (SR) are defined according to the strength of the MET cut, with the strongest being $> 700~\GEV$. In addition, the ratio of the leading jet $p_T$ to MET has to be larger than 0.5 to ensure the final state to be mono-jet-like. Strictly speaking, the event topology we are interested in involves two quark partons from the gauge boson decay. However, due to the highly boosted nature of the gauge boson, many events are tagged as a single collimated jet, and this analysis can be sensitive to probing the boosted gauge boson. 

\subsubsection*{Fat-jet Search}
In Ref.~\cite{Aad:2013oja}, 
\footnote{Very recently the CMS collaboration has also published results on searching for events with boson-tagged jet plus MET~\cite{CMS:2015jha}.}
a hadronically decaying gauge boson accompanied with large MET is looked for by reconstructing a large-radius jet (fat-jet). The jet is reconstructed with the Cambridge-Aachen algorithm with $R=1.2$~\cite{hep-ph/9707323}. A mass-drop filtering procedure is applied to identify the substructure of the large-radius jet~\cite{0802.2470}. The two leading subjets have to satisfy $\sqrt{y} > 0.4$, where 
\begin{eqnarray}
\sqrt{y} = \frac{{\rm min}(p_{T1},p_{T2})\Delta R}{m_{\rm jet}},
\end{eqnarray}
and $\Delta R = \sqrt{(\Delta \phi_{1,2})^2+(\Delta \eta_{1,2})^2}$, and $m_{\rm jet}$ is the mass of the large-radius jet. The large-radius jet is required to have $50~\GEV < m_{\rm jet} < 120~\GEV$, such that it is supposed to capture the hadronic $W$ and $Z$. The large-radius jet is further required to have $p_T > 250~\GEV$.

In addition, narrow jet is defined using the anti-$k_T$ jet algorithm, with $R=0.4$. Events with more than one narrow jet carrying $p_T > 40~\GEV$ and separated from the leading fat-jet $\Delta R > 0.9$, or separated from the MET with $\Delta \phi < 0.4$ are rejected. Two SRs are defined according to the MET threshold: $350~\GEV$ and $500~\GEV$ respectively.

\section{Interpretation as Resonance Search} \label{sec:RECAST}
In this Section, we impose the constraints on the 2 TeV resonance by recasting reported results of searches for events with a jet (boson-tagged or not) plus large MET. Concerning the optimization procedures, we extrapolate the current analyses by defining new SRs with higher MET cuts in order to derive stronger bounds. 
  
\subsection{Simulation Setup}
For both signal and background estimations, we have used the programs 
{\tt  MADGRAPH 5 v2.1.2}~\cite{Alwall:2014hca,Alwall:2011uj} interfaced to {\tt Pythia 6.4}~\cite{Sjostrand:2006za} and {\tt Delphes 3}~\cite{deFavereau:2013fsa} (which has {\tt FastJet} incorporated~\cite{Cacciari:2011ma,Cacciari:2005hq}).
 
%fat jet...... 
 
\subsubsection*{New Physics Signal}
As a benchmark model for the $X \to ZZ$ and $W^+W^-$ channels, we adopt the Randall-Sundrum (RS) graviton~\cite{Randall:1999ee}, which is a spin 2 boson and couples to the SM particle through the energy-momentum tensor.
For the case of  $X \to WZ$, we assume a $W'$ boson, which couples to SM particles in a similar way to the SM $W$ boson.
In both cases, we assume that the total decay width of the resonance is 100 GeV.
\subsubsection*{SM Background}
The SM background for SRs of interest is dominated by $Z$+jets, $W$+jets, diboson and top events.  
In order to study the prospect of the event searches' optimizations,
we need to estimate the number of the background events in optimized SRs characterized by the stronger MET cuts.
We first generate SM Monte-Carlo events and normalize the total number of events to the one in the ATLAS SRs. 
For both analyses, the normalization factor is consistent within a few tens of percent, compared to the cross section estimated with {\tt MADGRAPH}.

\subsection{Result}
\begin{figure}[h!]
\centering
\subcaptionbox{Mono-jet}{
\includegraphics[clip, width = 0.45 \textwidth]{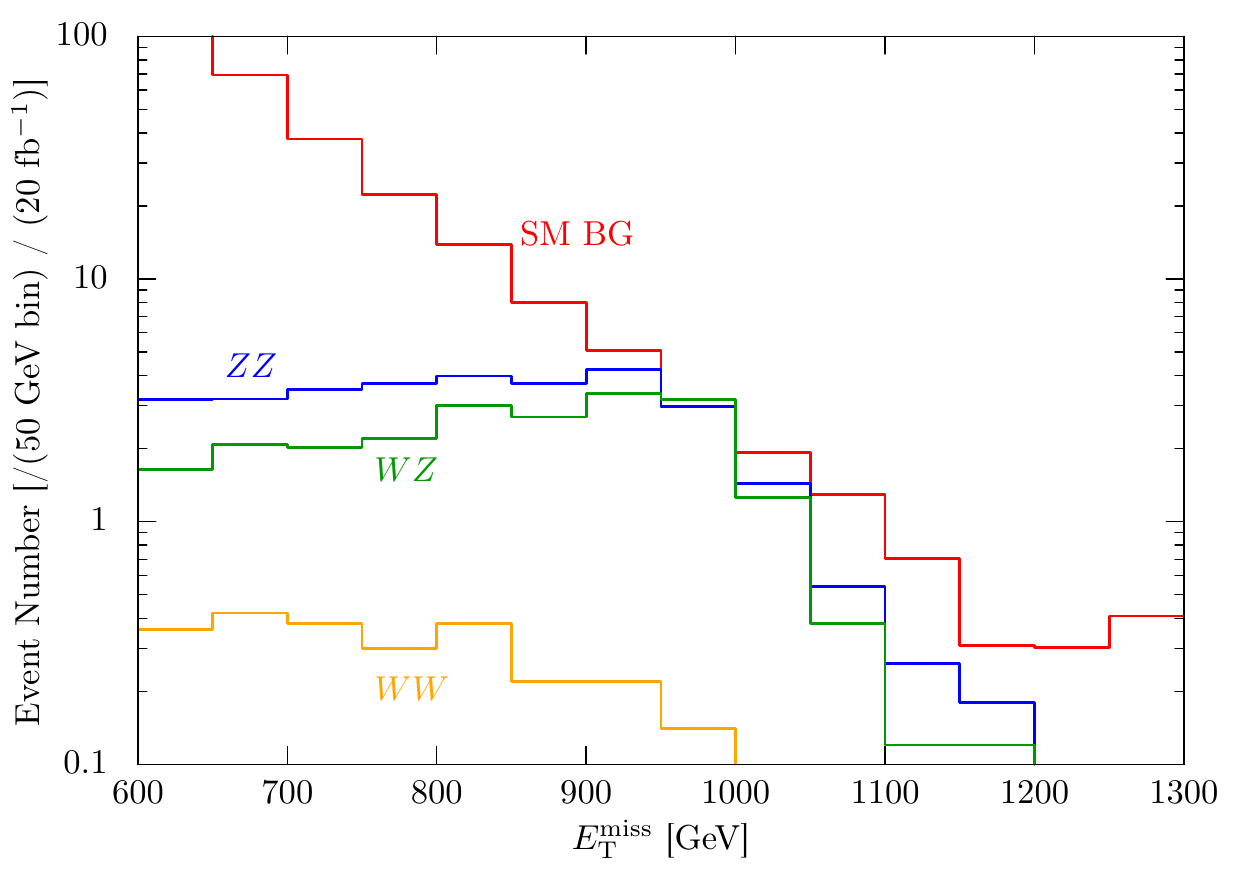}
}
\subcaptionbox{Fat-jet}{
\includegraphics[clip, width = 0.45 \textwidth]{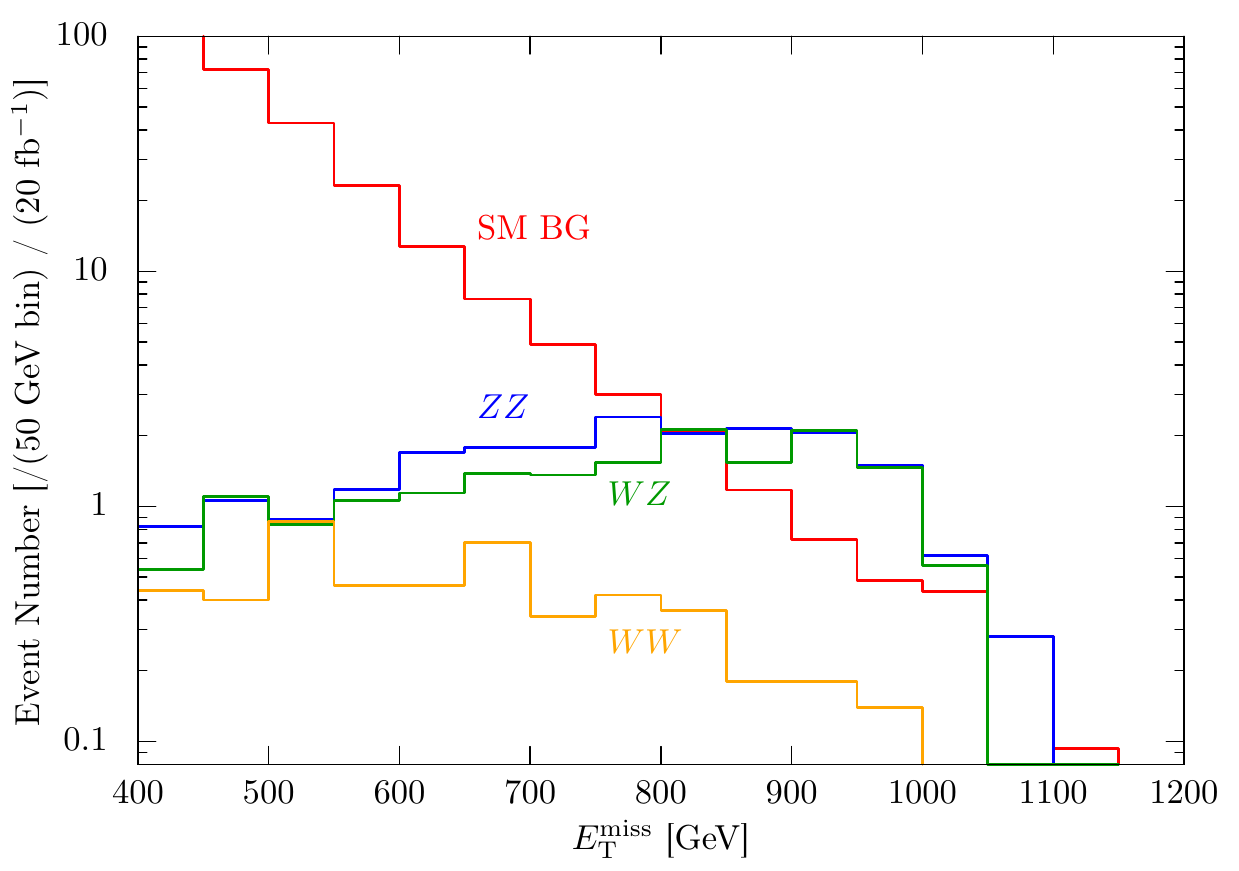}
}
\caption{Estimated missing transverse energy distribution of the SM background, the 2 TeV $ZZ$, $WZ$ and $WW$ diboson signal channels in the 8 TeV (a) mono-jet and (b) fat-jet analyses. The production cross section of the 2 TeV resonance has been normalized to 10 fb. Note the significant drop in the signal event numbers for bins with $E^{\rm miss}_{\rm T}\gtrsim 1$ TeV.
\label{fig:met}}
\end{figure}

We show the estimated missing transverse energy distribution of the SM background in Fig.~\ref{fig:met} at $\sqrt{s} = 8$ TeV.
We also show the missing transverse energy distribution in the cases that 
$m_X=2$ TeV and $\sigma_{pp\to X}=10$ fb.
The SM background decreases rapidly as the missing transverse energy is increased, while the signal distribution is relatively flat up to $E^{\rm miss}_{\rm T} \simeq 1$ TeV.
Above this threshold, the signal numbers are drastically reduced for $m_X = 2$ TeV is assumed.
Choices of the MET cut near this threshold can significantly enhance the signal to background ratio and achieve  strong constraints.

In Table~\ref{tab:mono}, we show the current status and prospect of the jet plus MET search at 8 TeV.
For reference, we show the observed and expected cross section times branching ratio upper limits on the 2 TeV resonances, with SM background-only hypothesis. In the following, a systematic uncertainty of 10\% is assumed for the expected background events.
As expected, resonances involving the decay into a $Z$ boson are well constrained.
Even though, at present, the observed cross section times branching ratio upper limits are slightly worse than the expected ones, no significant excesses from the SM background is observed. One can see the the upper limits can be significantly improved by increasing the MET cuts. 

Next, in Table~\ref{tab:fat}, we show the results of the 8 TeV fat-jet plus MET analysis. Again, no significant excess over the SM background is observed.
Compared to the previous mono-jet search, the background is reduced by about 90\% while the signal rate is reduced only by approximately 50\%.
Thanks to the large reduction of the SM background, the sensitivity is hugely improved.
Particularly, the MET cut 800 GeV will provide best sensitivity for the 2 TeV resonance searches. It is possible to impose an upper limit as low as 6 ${\rm fb}$ with this SR. 

 \begin{table}[htbp]
 \caption{\label{tab:mono} 
 Status and prospect for mono-jet search at 8 TeV. The observed number and expected SM background for MET cuts $\leq$ 700 GeV are taken from Ref.~\cite{Aad:2015zva}. For larger MET cuts, we extrapolate the ATLAS analysis, using our simulation results. A systematic uncertainty of 10\% is assumed for the background in the extrapolated SRs. Acceptance is calculated assuming a resonance of mass 2 TeV. The expected and observed cross section times branching ratio upper bound are given using $\mathrm{CL_s}$ procedure.
 }
\begin{tabular}{|c|c|rrrrr||rrr|}
\hline
\multicolumn{2}{|c|}{MET cut [GeV]} & $>350$ & $>400$ & $>500$ & $>600$ & $>700$ & $>800$ & $>900$ & $>1000$ \\
\cline{1-10}
\multicolumn{2}{|c|}{Observed Number} & $7988$ & $3813$ & $1028$ & $318$ & $126$ & -  & -& - \\
\hline
\multicolumn{2}{|c|}{SM BG} & $8300(300)$ & $4000(160)$ & $1030(60)$ & $310(30)$ & $97(14)$ & $36$  & $14$ & $5.4$  \\
\hline
\hline
&
Acceptance  &$0.20$ & $0.19$ & $0.18$ & $0.15$ & $0.12$  & $0.09$ & $0.06$
& $0.02$ \\
\cline{2-10}
$ZZ$& $\sigma^{\rm obs}_{95\%}$ [fb] &110  & 61  &38  &  25 & $24$  &  -   & - &- \\
\cline{2-10}
&
$\sigma^{\rm exp}_{95\%}$ [fb] &150  & 88  & 37 &  23 & $14$  & $8.8$ & $8.6$ & 16 \\
\hline \hline
& Acceptance & 0.13 &0.13& 0.12& 0.11& 0.10 & 0.08 & 0.05 & 0.01   \\
\cline{2-10}
$WZ$& $\sigma^{\rm obs}_{95\%}$ [fb] &170& 92 & 54 & 33 & 31 &  - & - & -  \\
\cline{2-10}
& $\sigma^{\rm exp}_{95\%}$ [fb] & 230 & 131 & 53 & 31 & 18 & 11 & 10 & 21   \\
\hline 
\hline
& Acceptance & 0.02 & 0.02 & 0.02 & 0.01 & 0.01& 0.01 & 0.004 & 0     \\
\cline{2-10}
$WW$& $\sigma^{\rm obs}_{95\%}$ [fb] &980& 560 & 400 & 290 & 310 & - & - & -  \\
\cline{2-10}
& $\sigma^{\rm exp}_{95\%}$ [fb] & 1400 & 820 & 400 &270 & 180 & 110 & 120 & -   \\
\hline
\end{tabular}
 \end{table}

\begin{table}[h]
 \caption{\label{tab:fat}  Status and prospect for fat-jet plus MET search at 8 TeV. The observed number and expected SM background for MET cuts $\leq$ 500 GeV are taken from Ref.~\cite{Aad:2013oja}. For larger MET cuts, we extrapolate the ATLAS analysis, using our simulation results. A systematic uncertainty of 10\% is assumed for the background in the extrapolated SRs. Acceptance is calculated assuming a resonance of mass 2 TeV. The expected and observed upper bound is given using $\mathrm{CL_s}$ procedure.  
 }
 \centering
\begin{tabular}{|c|c|rr||rrrrr|}
\hline
\multicolumn{2}{|c|}{MET cut [GeV]} & $>350$ &  $>500$ & $>600$ & $>700$ & $>800$ & $>900$ & $>1000$ \\
\hline
\multicolumn{2}{|c|}{Observed Number} & $705$ & $89$ & - &-  &-  &- &-    \\
\hline
\multicolumn{2}{|c|}{SM BG} & $707^{+48}_{-38}$ & $89^{+9}_{-12}$ & 30  & 12 & 4.7 & 1.8 & 0.7 \\
\hline
\hline
&Acceptance  &0.10& 0.09 & 0.08 & 0.06 & 0.05 & 0.03 & 0.007   \\
\cline{2-9}
$ZZ$& $\sigma^{\rm obs}_{95\%}$ [fb] &  54 & 18  & -  & - &- &- &- \\
\cline{2-9}
& $\sigma^{\rm exp}_{95\%}$ [fb] &  54 & 18 & 11 & 7.2 & 6.0 & 7.3 & 22 \\
\hline
\hline
&Acceptance  &0.09& 0.08 & 0.07& 0.06 & 0.04 & 0.03 & 0.007   \\
\cline{2-9}
$WZ$&$\sigma^{\rm obs}_{95\%}$ [fb] &  60 & 20  & -  & - &- &- &- \\
\cline{2-9}
&$\sigma^{\rm exp}_{95\%}$ [fb] &  61 & 20 & 12 & 7.5 & 6.6 & 7.5 & 22 \\
\hline
\hline
&Acceptance  &0.03& 0.02 & 0.02& 0.01 & 0.007 & 0.003 & 0   \\
\cline{2-9}
$WW$&$\sigma^{\rm obs}_{95\%}$ [fb] &  169 & 69  & -  & - &- &- &- \\
\cline{2-9}
&$\sigma^{\rm exp}_{95\%}$ [fb] &  173 & 68 & 53 & 44 & 44 & 56 & - \\
\hline
\end{tabular}
 \end{table}

\begin{figure}[h!]
\centering
\subcaptionbox{ $X\to ZZ$ \label{fig:cons_zz}}{\includegraphics[clip, width = 0.45 \textwidth]{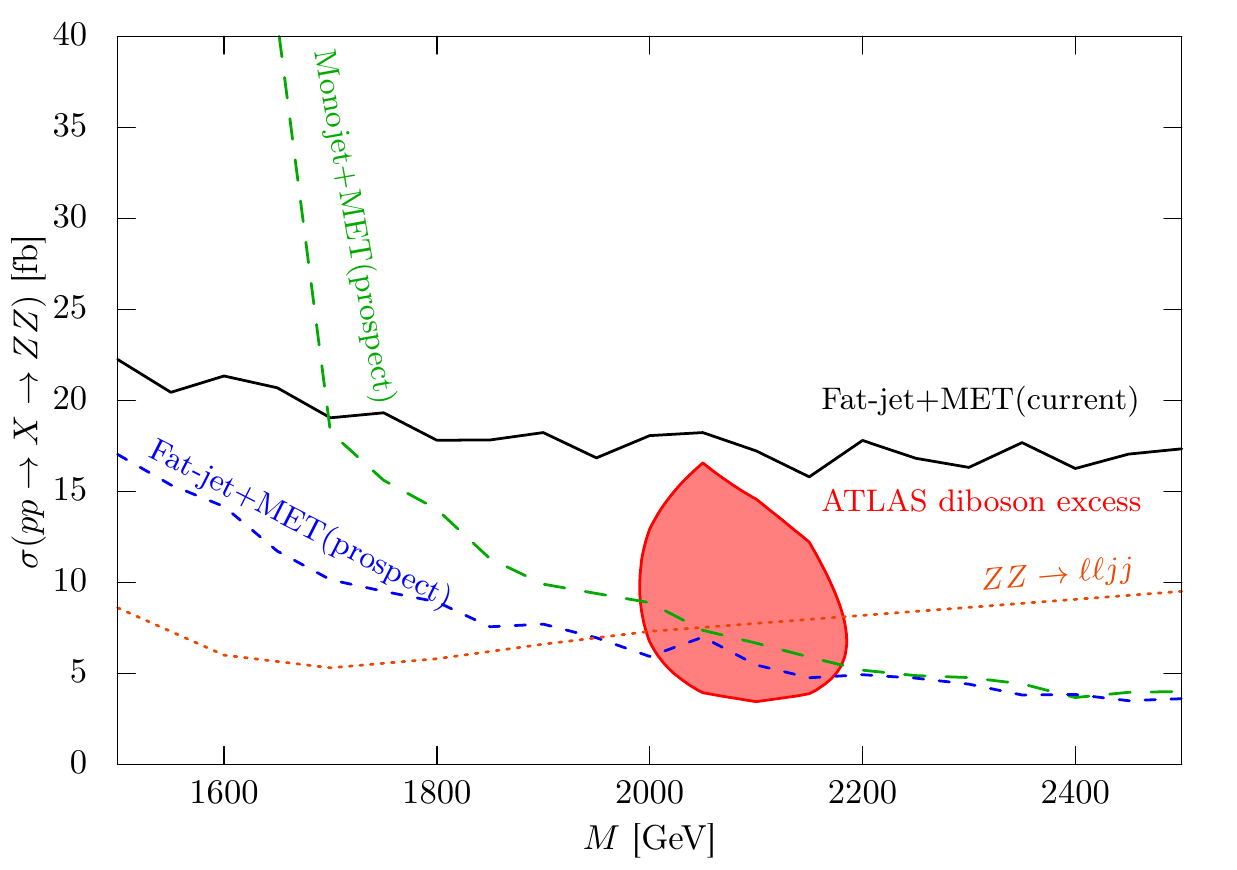}
}
\subcaptionbox{$X\to WZ$ \label{fig:cons_wz}
}{\includegraphics[clip, width = 0.45 \textwidth]{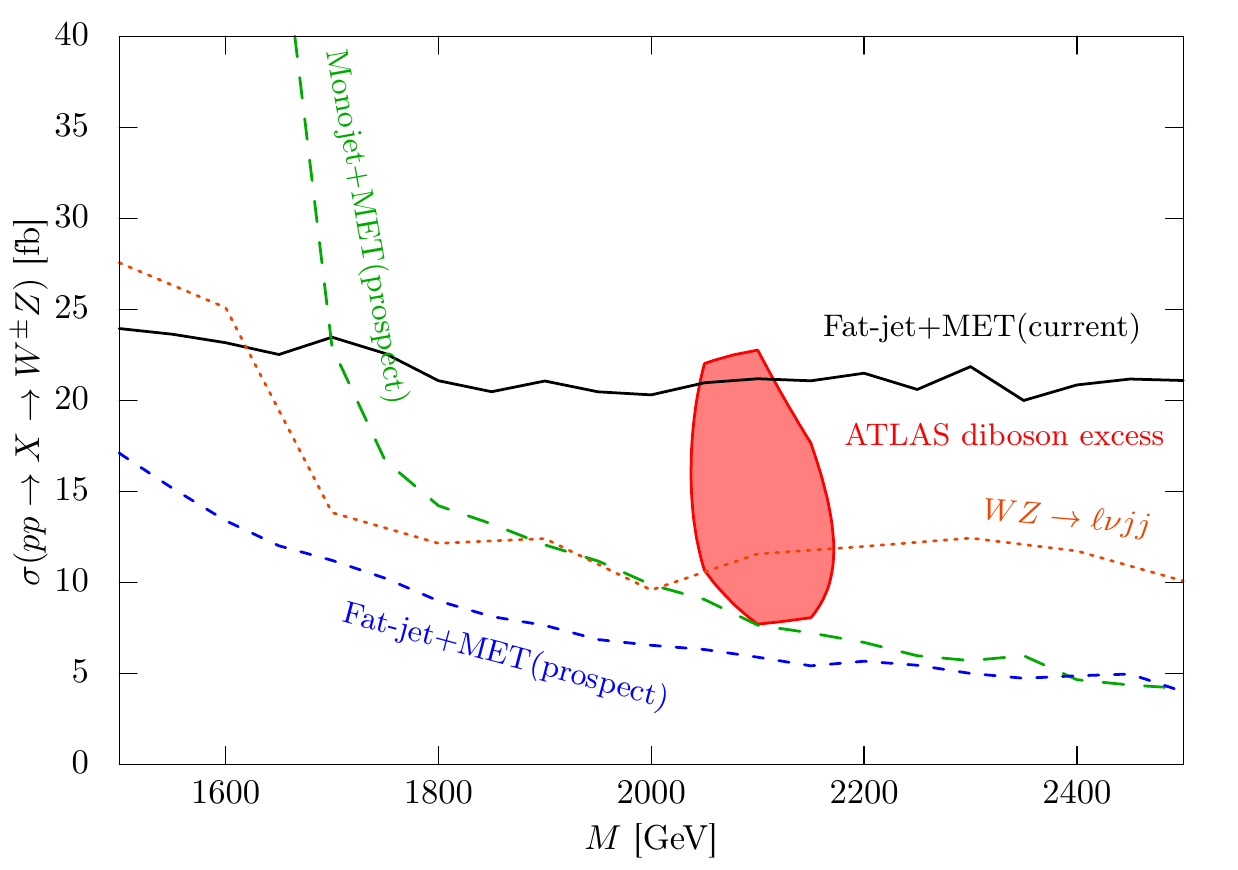}
}
\caption{\label{fig:cons} Current and prospective 8 TeV constraints on the diboson resonances ($X\to ZZ/WZ$). The shaded region is the $1\sigma$ favored region for explaining the observed excess. The black solid lines are the fat jet plus MET constraints derived with the current ATLAS results~\cite{Aad:2013oja}. The short-dashed blue lines show the optimized fat jet plus MET constraints, while the long-dashed green lines are the optimized jet plus MET constraints~\cite{Aad:2015zva}. The dotted lines represent constraints from the semi-leptonic diboson search~\cite{Aad:2014xka,Aad:2015ufa}.}
\end{figure}

In Fig. \ref{fig:cons}, we show the current and prospective constraints on the resonance mass-production cross section plane for the cases $X\to ZZ$ and $ZW$.
The red shaded regions are preferred region ($1\sigma$) to explain the ATLAS diboson excess.  
Here we assume the SM background distribution form $p_1 (1 - x)^{p_2} x^{p_3}$ with $x=m_{JJ}/\sqrt{s}$ and fit the signal distribution over the background in the dijet mass region of 1050-3550 GeV. Note that the resonance width is fixed to 100 GeV throughout the analysis. Changing the resonance width potentially shifts the $1\sigma$ region, i.e. a narrower resonance would shift the red shaded region towards $M = 2000~\GEV$ in Fig. \ref{fig:cons}. 
The solid black lines show the current observed cross section times branching ratio upper bound with the fat-jet plus MET search.
The long-dashed green (short-dashed blue) lines represent prospects for mono-jet (fat-jet plus MET) searches.
In Fig. \ref{fig:cons_zz}, we also show the constraints with $X\to Z(\to \ell^+ \ell^-) + Z(\to 2j)$ \cite{Aad:2014xka} for reference.
Note that in Ref. \cite{Aad:2014xka}, the constraint for the region $m_X>2$ TeV is not given. We have extrapolated the constraint assuming the reconstruction efficiency for the ``merged region"  to be 30\%.
The dotted line in Fig. \ref{fig:cons_wz} shows the constraint with $X\to W(\to \ell\nu) + Z(\to 2j)$ \cite{Aad:2015ufa}. The current jet plus MET constraints are relatively weak and are not shown in the figure. As can be observed from the figure, even though the current fat-jet plus MET analysis does not constrain the parameter region favoring the excess, a simple optimization MET cut can greatly improve the bound, and a large portion of parameter region can potentially be tested.

Finally, let us note that in our case, the plausible value of MET is around 1 TeV at maximum as the resonance mass is approximately 2 TeV. Raising the MET cut above 1 TeV would not hugely improve the constraints on the observed 2 TeV excess, as shown explicitly in Fig.~\ref{fig:met}. This statement holds true even for the LHC Run 2 prospect.

\section{LHC Run 2 Prospect} \label{sec:13}
In this section, we study projected sensitivities of the mono-jet and fat-jet searches on testing the diboson resonance at 13 TeV. We assume an integrated luminosity of 10 ${\rm fb}^{-1}$ and 100 ${\rm fb}^{-1}$. The number of background events are estimated by using {\tt MADGRAPH}. As compared to the 8 TeV analysis, the only optimization that we have performed is by changing the MET cut to larger values. Further improvements could be achieved by tuning other selection cuts, but an accurate description is only feasible after the 13 TeV data is collected and analyzed. The projection results presented here are therefore conservative.

\subsection{Signal Cross Section}
\begin{figure}[h!]
\centering
\includegraphics[clip, width = 0.5 \textwidth]{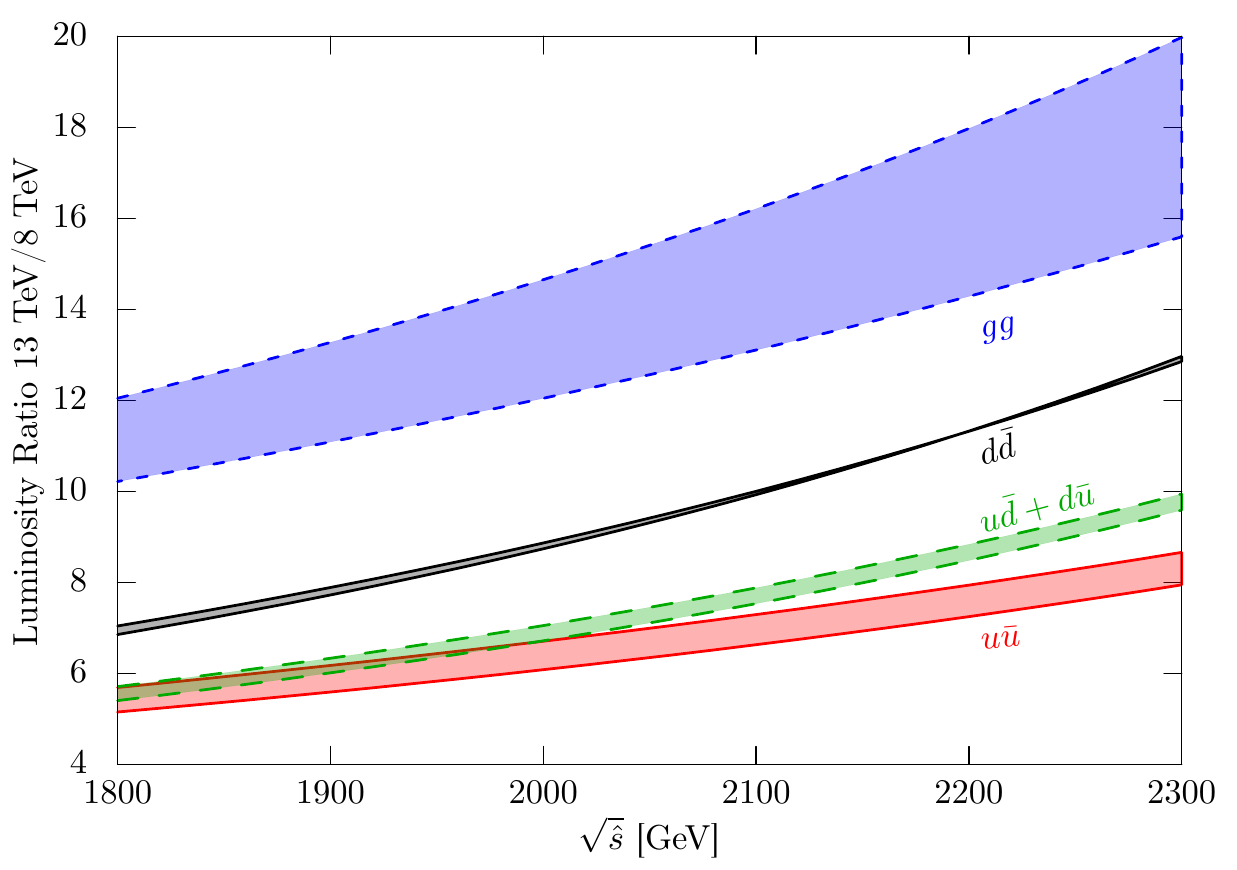}
\caption{Parton luminosity ratios 13/8 TeV.
\label{fig:ratio}}
\end{figure}

In order to explain the ATLAS diboson excess, the plausible cross section of 
a pair of the weak bosons mediated by the 2 TeV resonance is around 10 fb for $\sqrt{s}=8$ TeV.
In order to study the prospect at the 13 TeV running LHC, we need to estimate the relevant cross section.
The cross section depends on details of the new physics model. Let us consider the case where the production channel of the resonance $X$ is dominated by $PP'$ parton-level collision, i.e. $PP' \to X$.
When the narrow width approximation is assumed, the relation between the cross sections at 8 TeV and 13 TeV can be approximately given by
\begin{align}
\frac{\sigma_{pp\to X} |^{PP'}_{\sqrt{s}=13{\rm TeV}}}{\sigma_{pp\to X} |^{PP'}_{\sqrt{s}=8{\rm TeV}}}
\simeq
\left. 
\frac{{\cal L}^{PP'}_{\sqrt{s}=13{\rm TeV}}}{{\cal L}^{PP'}_{\sqrt{s}=8{\rm TeV}}}
\right|_{\sqrt{\hat{s}}=m_X},
\end{align}
where ${\cal L}^{PP'}$ is the parton luminosity of the partons $P$ and $P'$.
With these specification, we can estimate the 13 TeV production cross section with the parton luminosity ratio between 8 TeV and 13 TeV.
This ratio is shown in Fig.~\ref{fig:ratio}.
Here we use CTEQ6L \cite{Stump:2003yu} and  MSTW2008LO \cite{Martin:2009iq} PDF sets to indicate the uncertainty of the luminosity ratio.
It is seen that for instance, $W'$ dominantly comes from $u\bar{d}$ partons, and the enhancement of the cross section at 13 TeV is around 6.
The gluon fusion channel gives the largest 13 to 8 TeV enhancement of the cross section.
The $u\bar{u}$ fusion channel performs worst, giving an enhancement factor of about 6.

\subsection{Sensitivity at 13 TeV}
As in the 8 TeV analyses, we use the RS graviton ($W'$ boson) as our benchmark model to test the $ZZ,WW~(WZ)$ diboson channel.
In Tables~\ref{tab:monoj13} and~\ref{tab:monof13}, we show the expected production cross section times branching ratio upper limits of mono-jet and fat-jet plus MET searches on the 2 TeV with several choices of MET cut up to 1 TeV.

As discussed previously, depending on the dominant production cross section, the signal cross section can be enhanced up to a factor of ten compared to the 8 TeV run. It is however noted that the SM background events are also similarly enhanced by several factors.
Therefore, compared to the 8 TeV prospect, the constraint is not so drastically improved, particularly if the resonance production mainly comes from valence quark fusion.
If the dominant production is via the gluon fusion channel, as in the case of RS graviton, the obtained constraints are  effectively twice stronger at 10 fb$^{-1}$. 

Further optimization of the MET cut (e.g., $>$ a few  TeV) would not greatly improve the run 2 reach of the 2 TeV resonance. 
This is because, as discussed in the previous section, it is kinematically implausible to have MET larger than 1 TeV originating from a 2 TeV resonance. While a stronger MET cut would lead to a reduction of SM background, the observed signal events suffer from a more drastic drop in number. Hence, the 13 TeV projections performed in this paper (with MET cuts at around 1 TeV) are considered to be optimal and realistic.

Let us comment on uncertainties of the present prospects.
Our background estimation is just based on Monte-Carlo simulation with  the fast detector simulation.
For the accurate estimation, we need full detector simulation and real data as control regions.
We expect our estimation will suffer from several tens of percent uncertainty.
In the case of the low-background, i.e., tight cuts and the lower integrated luminosity, this uncertainty is not so significant for the  estimation of the reach of $X$ searches, since it is determined mainly by statistical fluctuations of the background and signal. 
However, if the background number gets larger, this uncertainty directly affects the prospects.
Moreover, in such cases, detailed estimation of the  systematic uncertainty, which we have assumed 10 \%/, is crucially important.
This estimation needs more detailed full detector simulation.

%As in the 8 TeV analyses, we use the RS graviton ($W'$ boson) as our benchmark model to test the $ZZ~(WZ)$ diboson channel. An excess corresponding to a total cross section of $\sim20$ fb has been observed at resonance of mass around 2 TeV. Extrapolating the production cross section using MADGRAPH, we estimate that the excess corresponds to a total production cross section of $\sim 4$ pb at 13 TeV. In Tables~\ref{tab:monoj13} and~\ref{tab:monof13}, we show the production cross section upper limits of mono-jet and fat-jet searches on the 2 TeV with several choices of MET cut up to around 1 TeV. It can be seen that production cross section up to $O(10)$ fb can be excluded at 95\% confidence level, providing a rigorous test on the validity of the excess.

 \begin{table}[htbp]
 \centering
 \caption{Prospect for jet plus MET search at 13 TeV with integrated luminosities of 10 and 100 ${\rm fb}^{-1}$. The SM background production cross section is estimated using {\tt MADGRAPH}. A systematic uncertainty of 10\% is assumed for the background. Acceptance is calculated assuming a resonance of mass 2 TeV. The expected and observed upper bound is given using $\mathrm{CL_s}$ procedure.}
  \label{tab:monoj13}
\begin{tabular}{|c|c|rrrrrr|}
\hline
\multicolumn{2}{|c|}{MET cut [GeV]}  & $>500$ & $>600$ & $>700$ & $>800$ & $>900$ & $>1000$ \\
\hline
\multicolumn{2}{|c|}{SM BG [fb]} & 180 & 70 & 30 & 13 & 6 & 3 \\
\hline
\hline
&Acceptance   &0.15 & 0.13 & 0.11 & 0.08 & 0.05 & 0.02  \\
\cline{2-8}
$ZZ$ &$\sigma^{\rm exp}_{95\%}$ @ $10~{\rm fb}^{-1}$  [fb] &230& 110 & 63 & 44 & 43 & 79  \\
\cline{2-8}
 &$\sigma^{\rm exp}_{95\%}$ @ $100~{\rm fb}^{-1}$  [fb] &220& 100 & 54 & 34 & 27 & 38  \\
\hline
\hline
&Acceptance   &0.12 & 0.11 & 0.09 & 0.08 & 0.05 & 0.02  \\
\cline{2-8}
$ZW$ &$\sigma^{\rm exp}_{95\%}$ @ $10~{\rm fb}^{-1}$  [fb] &310 & 130 & 70 & 47 & 47 & 78  \\
\cline{2-8}
 &$\sigma^{\rm exp}_{95\%}$ @ $100~{\rm fb}^{-1}$  [fb] &300 & 120 & 61 & 36 & 30 & 37  \\
\hline
\hline
&Acceptance   &0.01 & 0.01 & 0.007 & 0.005 & 0.002 & 0  \\
\cline{2-8}
$WW$ &$\sigma^{\rm exp}_{95\%}$ @ $10~{\rm fb}^{-1}$  [fb] &2800 & 1400 & 890 & 740 & 973 & -  \\
\cline{2-8}
 &$\sigma^{\rm exp}_{95\%}$ @ $100~{\rm fb}^{-1}$  [fb] &2700 & 1400 & 800 & 550 & 580 & -  \\
\hline
\end{tabular}
 \end{table}
 
\begin{table}[h]
\centering
 \caption{Prospect for fat-jet plus MET search at 13 TeV with integrated luminosities of 10 and 100 ${\rm fb}^{-1}$. The SM background production cross section is estimated using {\tt MADGRAPH}. A systematic uncertainty of 10\% is assumed for the background. Acceptance is calculated assuming a resonance of mass 2 TeV. The expected and observed upper bound is given using $\mathrm{CL_s}$ procedure.}
   \label{tab:monof13}
\begin{tabular}{|c|c|rrrrrr|}
\hline
\multicolumn{2}{|c|}{MET cut [GeV]} & $>500$ & $>600$ & $>700$ & $>800$ & $>900$& $>1000$  \\
\hline
\multicolumn{2}{|c|}{SM BG [fb]} & 25 & 10 & 4.5& 2.1 & 1.0 & 0.5 \\
\hline
\hline
%$N_{\rm inc}$ & $25.2$ & $17.7$  & $12.5$& $9.3$ \\
%\hline
&Acceptance  &0.07&0.06&0.05&0.04&0.02&0.006 \\
\cline{2-8}
$ZZ$ &$\sigma^{\rm exp}_{95\%}$ @ $10~{\rm fb}^{-1}$ [fb]&82&49&36&30&36&83  \\
\cline{2-8}
&$\sigma^{\rm exp}_{95\%}$ @ $100~{\rm fb}^{-1}$ [fb] 
&70&34&20&14&14&28 \\
\hline
\hline
 &Acceptance  &0.06 & 0.06 & 0.05 & 0.04 & 0.02 & 0.01 \\
\cline{2-8}
$ZW$ &$\sigma^{\rm exp}_{95\%}$  @ $10~{\rm fb}^{-1}$ [fb] &96 & 52 & 37 & 29 & 34 & 67  \\
\cline{2-8}
 &$\sigma^{\rm exp}_{95\%}$  @ $100~{\rm fb}^{-1}$ [fb] &81 & 37 & 20 & 14 & 13 & 22  \\
\hline
\hline
&Acceptance  &0.01 & 0.01 & 0.007 & 0.004 & 0.002 & 0 \\
\cline{2-8}
$WW$ &$\sigma^{\rm exp}_{95\%}$ @ $10~{\rm fb}^{-1}$ [fb] &430 & 290 & 270 & 270 & 410 & -  \\
\cline{2-8}
 &$\sigma^{\rm exp}_{95\%}$ @ $100~{\rm fb}^{-1}$ [fb] &360 & 200 & 150 & 130 & 150 & -  \\
\hline
\end{tabular}

 \end{table}

\section{Summary and Discussion}\label{sec:summary}
In this paper, we have discussed the current constraints and prospect for the jet and fat-jet plus  MET signal for testing the ATLAS diboson excess.
We found that the $V(\to 2j)Z(\to\bar{\nu}\nu)$ mode is one of the most powerful probes of the diboson excesses, and a large portion of the favored region can be tested even with the LHC Run 1 data.
Simple optimizations of the MET cut can greatly enhance the sensitivity.
If the observed excess is due to new physics-initiated dibosons, we expect to observe excesses in this channel.

This channel provides a strong discriminant, i.e., the channel is very sensitive to the case that the decay product of the heavy resonance includes a $Z$ boson and less sensitive to the $X\to W^+W^-$ channel.
Therefore, this channel can provide us information on the decay mode of the resonance $X$.
By combining the searches for the other decay channel, we will get insights into the details of the nature of $X$.

We have thus far focused on analyzing the decay products of the weak bosons, which one expects to be model independent. Let us comment on possible model-dependent effects. 
Depending on the production channel and the model parameter, the momentum distribution of the partons can differ and may influence the cut efficiencies.
Moreover, initial state radiation, of which the signal rate is model dependent, can affect the jet-veto cuts employed in the dark matter searches.
Therefore, in addition to the simple MET optimizations performed in this paper, further tunings to improve the resonance search is also possible, which would lead to a more powerful probe of the resonance.

%%%%%%%%%%%%%%%%%%%%%%%%%%%%%%%%%%%%
\section*{Acknowledgments}
%%%%%%%%%%%%%%%%%%%%%%%%%%%%%%%%%%%%
We thank M. Stoll for providing us useful information about the fat-jet analysis, and T. T. Yanagida for encouragement leading to this work. We are also grateful to K. Hamaguchi for commenting on the manuscript.
SPL is supported by JSPS Research Fellowships for Young Scientists and the Program for Leading Graduate Schools, MEXT, Japan.

%%%%%%%%%%%%% References %%%%%%%%%%%%%%%%%%%
\bibliographystyle{aps}
\bibliography{ref}
%%%%%%%%%%%%%%%%%%%%%%%%%%%%%%%%%%%%%%%%%%%%

\end{document}